\documentclass[twocolumn]{aastex631}

\usepackage{amsmath}

\usepackage[english]{babel}
\usepackage[autostyle]{csquotes}

\begin{document}

\title{Observational Signatures of Dust Traffic Jams in Polar-aligning Circumbinary Disks}

\author{Jeremy L. Smallwood}
\affiliation{Institute of Astronomy and Astrophysics, Academia Sinica \\
Taipei 106216, R.O.C.\\
}

\author{Rebecca Nealon}
\affiliation{Centre for Exoplanets and Habitability, University of Warwick\\
Coventry CV4 7AL, UK\\
}
\affiliation{Department of Physics, University of Warwick\\
Coventry CV4 7AL, UK\\
}

\author{Hsi-Wei Yen}
\affiliation{Institute of Astronomy and Astrophysics, Academia Sinica \\
Taipei 106216, R.O.C.\\
}

\author{Christophe Pinte}
\affiliation{School of Physics and Astronomy, Monash University \\
Vic 3800, Australia \\
}
\affiliation{Universit\'e Grenoble Alpes, CNRS, IPAG\\
F-38000 Grenoble, France\\
}

\author{Cristiano Longarini}
\affiliation{Institute of Astronomy, University of Cambridge\\
Madingley Rd, CB3 0HA Cambridge, United Kingdom \\
}

\author{Hossam Aly}
\affiliation{Faculty of Aerospace Engineering, Delft University of Technology \\
Kluyverweg 1, 2629 HS Delft, The Netherlands\\
}

\author{Min-Kai Lin}
\affiliation{Institute of Astronomy and Astrophysics, Academia Sinica \\
Taipei 106216, R.O.C.\\
}



\begin{abstract}
Misaligned circumbinary disks will produce dust traffic jams during alignment or anti-alignment to the binary orbital plane. We conduct a hydrodynamical simulation of an initially misaligned circumbinary disk undergoing polar alignment with multiple dust species. Due to differential precession between the gas and dust components, multiple dust traffic jams are produced within the disk during polar alignment. The radial locations of the dust traffic jams depend on the Stokes number of the grains, which depends on grain size. We compute the dust temperature structure using post-processing radiative transfer to produce continuum images at cm-wavelengths. Multiple distinct rings emerge in the continuum images, corresponding to the dust traffic jams.  The angular resolution of upcoming observations from SKA and ngVLA will be sufficient to detect centimeter-sized grains in protoplanetary disks and resolve the widths of dust traffic jams. Therefore, dust traffic jams resulting from the differential precession of gas and dust in misaligned circumbinary disks will be a prime target for more extended wavelength observations.
\end{abstract}



\section{Introduction} 
\label{sec::intro}

In our galaxy, it is estimated that a significant portion, ranging from $40\%$ to $50\%$ of stars, form binary systems \citep{Duquennoy1991,Mayor2011,Raghavan2010,Tokovinin2014a,Tokovinin2014b}. Young binary star systems often house a circumbinary disk of gas and dust, providing a potential site for circumbinary planet formation. Observations consistently reveal that these circumbinary disks frequently show misalignment relative to the binary orbital plane \cite[e.g.,][]{Czekala2019}. A low-mass circumbinary disk with significant misalignment will precess around the binary eccentricity vector. This precession leads to the alignment of the angular momentum vector of the disk with the binary eccentricity vector, resulting in a polar-aligned circumbinary disk \citep{Aly2015,Martinlubow2017,Martin2018,Lubow2018,Zanazzi2018,MartinLubow2019}. 

Recently, \cite{Smallwood2024a,Smallwood2024b} investigated the polar alignment of circumbinary disks with gas and dust. Using hydrodynamical simulations, they found that dust traffic jams formed within circumbinary disks can undergo polar alignment due to the differential precession of the gas and dust \cite[e.g.,][]{Aly2020,Longarini2021,Aly2021}. The formation of dust traffic jams is robust for varying binary and disk parameters, which may be necessary to form circumbinary planets, precisely polar circumbinary planets. \cite{Smallwood2024b} used 2D shearing box calculations of streaming instability growth rates and growth timescales. They found unstable modes for dust Stokes number greater than unity. Moreover, it was observed that the growth timescale of the streaming instability is shorter than the oscillation timescale of the tilt during the polar alignment process. As a result, the dust ring is expected to be long-lived once the gas disk achieves polar alignment, indicating that the streaming instability may play an essential role in forming polar planets.

Observations have revealed two known polar circumbinary disks around the binary star systems, 99 Herculis (99 Her) \citep{Kennedy2012} and HD 98800B \citep{Kennedy2019}. 99 Her, a $\sim 10\, \rm Gyr$ old binary, houses the only known polar debris disk.  This configuration is plausible if the gas carries the dust to a polar orientation during the early stages of the binary lifetime \citep{Smallwood2020a}. The polar disk around HD 98800B shows traces of gas still present \citep{Kennedy2019}. Another nearly polar-aligned circumbinary disk was presumed around the post–asymptotic giant branch (AGB) star binary AC Her \citep{Martin2023}. Based on the disk structure, AC Her might have evidence of the first polar circumbinary planet \citep{Martin2023}, but subsequent observations are needed for confirmation. Circumbinary planets with polar orbits may form with comparable efficiency to coplanar planets \citep[e.g.,][]{Childs2021}.

This letter provides observational signatures of the dust traffic jams in polar-aligning circumbinary disks. Providing synthetic observations of misaligned circumbinary disks will help further constrain grain size distribution and grain growth in circumbinary disks. Misaligned circumbinary disks will be prime candidates for upcoming observations from ngVLA and SKA. We organize the letter as follows. In Section~\ref{sec::methods}, we showcase the setups for our hydrodynamical simulation and radiative transfer calculations. In Section~\ref{sec::Results}, we present the results of the hydrodynamical simulation. We use the results of the hydrodynamical simulations as input for the synthetic observations, which we show in Section~\ref{sec::syn_obs}. Last, we give a discussion and conclusion in Sections~\ref{sec::Discussion} and~\ref{sec::Conlusion}, respectively.


\section{Methods} 
\label{sec::methods}

\subsection{Hydrodynamics}
\label{sec::hydro_setup}
We use the two-fluid approximation \citep{Laibe2012a,Laibe2012b} in the 3-dimensional smoothed particle hydrodynamics code {\sc phantom} \citep{Price2018} to model an inclined circumbinary disk with gas and dust components.  A useful measure for describing the coupling between dust and gas in the Epstein regime is the Stokes number, which is defined as
\begin{equation}
    \rm St = \frac{\pi}{2} \frac{\rho_{\rm d}s}{\Sigma_{\rm g}},
    \label{eq::st}
\end{equation}
where $\rho_{\rm d}$ is the dust intrinsic density and $\Sigma_{\rm g}$ is the gas surface density. Given our focus on the polar alignment of a circumbinary disk characterized by low gas density and $\rm St > 1$, the two-fluid algorithm is a fitting choice for our simulation. This implementation incorporates drag heating effects but does not consider the thermal coupling between gas and dust, as detailed in \cite{Laibe2012a}.

We set up an equal-mass binary star system, with $M_1 = M_2 = 0.5\, \rm M_{\odot}$, where $M_1$ is the mass of the primary star and $M_2$ is the mass of the secondary star. The total binary mass is thus $M = M_1 + M_2$. The binary has an initial separation $a_{\rm b} = 16.5\, \rm au$ and eccentricity $e_{\rm b} = 0.8$, with a binary orbital period $\rm P_{orb} = 67\, \rm yr$ (resembling the 99 Her binary). We model the binary as a pair of sink particles with an accretion radius of $1.2a_{\rm b} = 20\, \rm au$. Particles crossing the accretion radius transfer their mass and angular momentum to the sink, treated as a rigid boundary. To expedite computations, the sink accretion radii are set to be similar to the binary separation, and particle orbits within the binary cavity are not resolved. We note that this choice of accretion radius does not impact our simulation results \cite[refer to Section 3.5 in][]{Smallwood2024a}. The simulation runs for $1000\, \rm P_{orb}$ or $\sim 67,000\, \rm yr$.

The circumbinary disk is initially modeled with $500,000$ equal-mass gas particles and $50,000$ dust particles. These particles are distributed within the inner disk radius, $r_{\rm in} = 40 \rm au$ or $2.5a_{\rm b}$, and the outer disk radius, $r_{\rm out} = 120 \rm au$ or $7a_{\rm b}$. The initial disk mass is set to $M_{\rm d} = 0.001M_{\odot}$, and the initial tilt is $i_0 = 60^\circ$. The minimum tilt for polar alignment is given by
\begin{multline}
i_{\rm min} = \arccos \\ \left[   \frac{\sqrt{5} e_{\rm b0} \sqrt{4e_{\rm b0}^2 - 4j_0^2(1 - e_{\rm b0}^2) + 1} - 2j_0(1 - e_{\rm b0}^2)}{1 + 4e_{\rm b0}^2} \right],
\end{multline}
as derived by \citet{Martin2019}. Here, $j_0 = J_{\rm d0}/J_{\rm b0}$ represents the initial angular momentum ratio of the disk to the binary, and $e_{\rm b0}$ denotes the initial binary eccentricity. The critical tilt for this system to align polar is $i_{\rm crit} \sim 38^\circ$. The gas surface density profile is initially modeled as a power-law distribution given by
 \begin{equation}
     \Sigma(r) = \Sigma_0 \bigg( \frac{r}{r_{\rm in}} \bigg)^{-p},
     \label{eq::sigma}
 \end{equation}
where $\Sigma_0 =  6\times10^{-3}\, \rm g/cm^2$ is the density normalization (based on the initial disk mass), $p$ is the power law index, and $r$ is the cylindrical radius 
We set $p=+3/2$. It is important to note that we disregard the influence of disk self-gravity because of the selected total initial disk mass. We employ a local isothermal equation-of-state (EOS), expressed as
\begin{equation}
c_{\rm s} = c_{\rm s,in} \left( \frac{r}{r_{\rm in}} \right)^{-q},
\end{equation}
where $c_{\rm s,in}$ represents the sound speed at the inner radius, derived from hydrostatic equilibrium. The disk thickness, denoted as $H$, exhibits a radial scaling given by
\begin{equation}
H = \frac{c_{\rm s}}{\Omega} \propto r^{3/2-q},
\end{equation}
where $\Omega = \sqrt{GM/r^3}$ and $q = +3/4$. The selected values of $p$ and $q$ gives (initial) uniform resolution in the disk, and hence uniform resultant physical viscosity. We initialize the gas disk with an aspect ratio of $H/r = 0.1$ at $r = r_{\rm in}$. The viscosity, $\alpha_{\rm SS}$, according to the \cite{Shakura1973} prescription, is defined as
\begin{equation}
\nu = \alpha_{\rm SS} c_{\rm s} H,
\end{equation}
where $\nu$ represents the kinematic viscosity. For the computation of $\alpha_{\rm SS}$, we apply the prescription from \cite{Artymowicz1994}, expressed as 
\begin{equation}
\alpha_{\rm SS} \approx \frac{\alpha_{\rm AV}}{10}\frac{\langle h \rangle}{H},
\end{equation}
where $\langle h \rangle$ is the average particle smoothing length. We set the \cite{Shakura1973} viscosity parameter as $\alpha_{\rm SS} = 0.01$, resulting in an artificial viscosity value of $\alpha_{\rm AV} = 2.4$. 
The viscosity prescription incorporates an additional parameter, $\beta_{\rm AV}$, serving as a non-linear term initially introduced to prevent particle penetration in high Mach number shocks \cite[e.g.,][]{Monaghan1989}, and is set to $\beta_{\rm AV} = 2.0$ \cite[e.g.,][]{Nealon2015}.

The dust particles are initially distributed following the same surface density profile as the gas, with a dust-to-gas mass ratio of $0.01$. The simulation includes four different dust species with a grain size power-law index of $3.5$.  Therefore, we simulate particle sizes of $s = 0.7, 1.2, 2.1, 3.7,\rm cm$, which correspond to Stokes numbers ranging from [$\sim 6$ to $\sim 27$], [$\sim 10$ to $\sim 49$], [$\sim 18$ to $\sim 84$], and [$\sim 31$ to $\sim 140$], respectively, from $r_{\rm in}$ to $r_{\rm out}$.  The Stokes number varies as a function of radius, but the dust particle size remains constant. We take the intrinsic grain density to be $3.00\, \rm g/cm^3$.  The initial dust disk aspect ratio is equivalent to the gas disk aspect ratio.

\begin{figure*} 
\includegraphics[width=2\columnwidth]{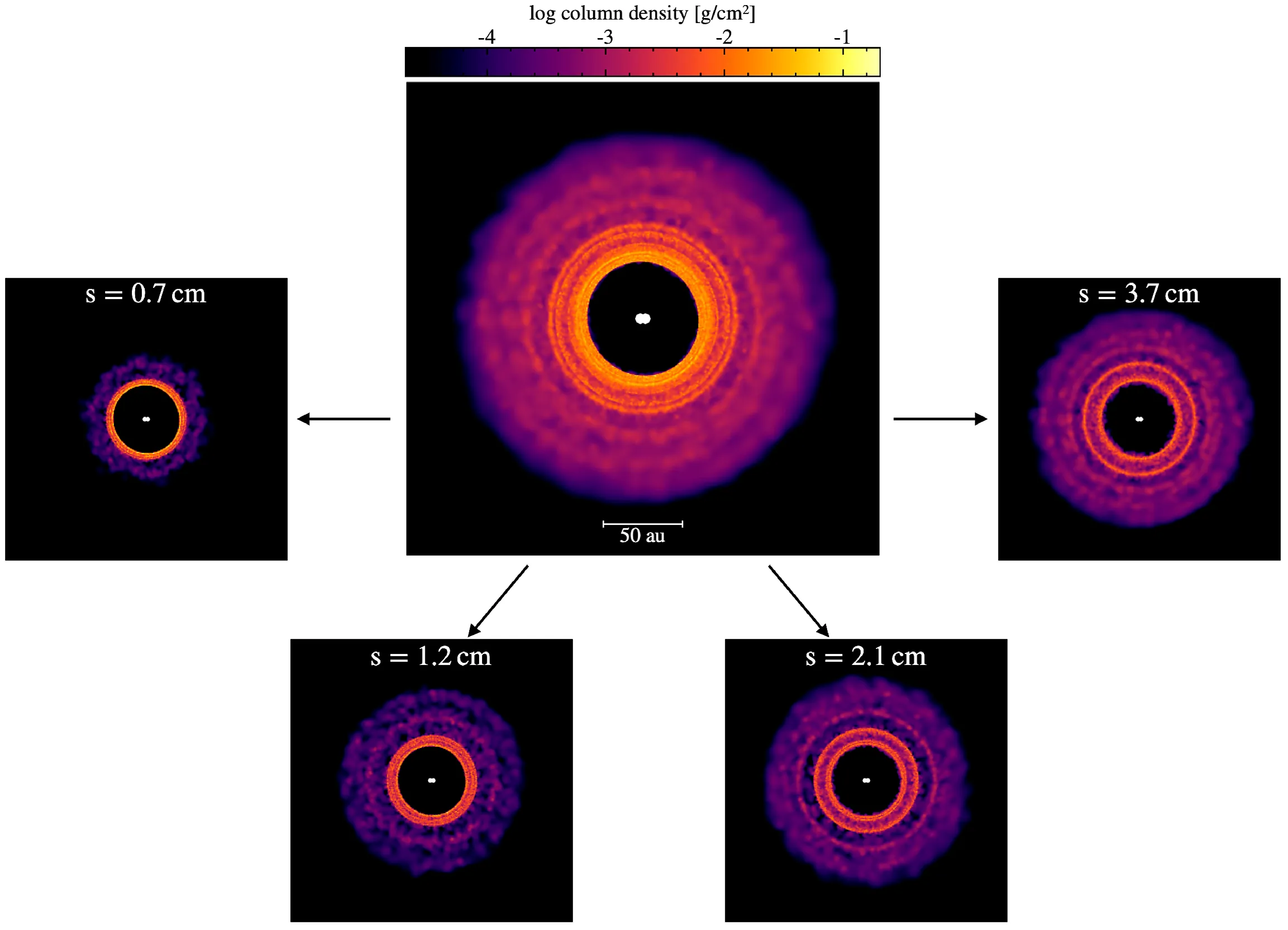}
\centering
\caption{The dust structure of a nearly polar dusty circumbinary disk with multiple dust species. The center image shows the circumbinary disk represented by all dust species. Each dust species is individually rendered and shown in the smaller panels: $\rm s = 0.7\, cm$,  $\rm s = 1.2\, cm$, $\rm s = 2.1\, cm$, and $\rm s = 3.7\, cm$. All panels view the disk in the $y$-$z$ plane at $1000\, \rm P_{orb}$. The color denotes the dust surface density. The differential precession between the gas and dust during polar alignment produces multiple dust rings with radial locations depending on the grain size.}
\label{fig::splash}
\end{figure*}

\begin{figure*} 
\includegraphics[width=2\columnwidth]{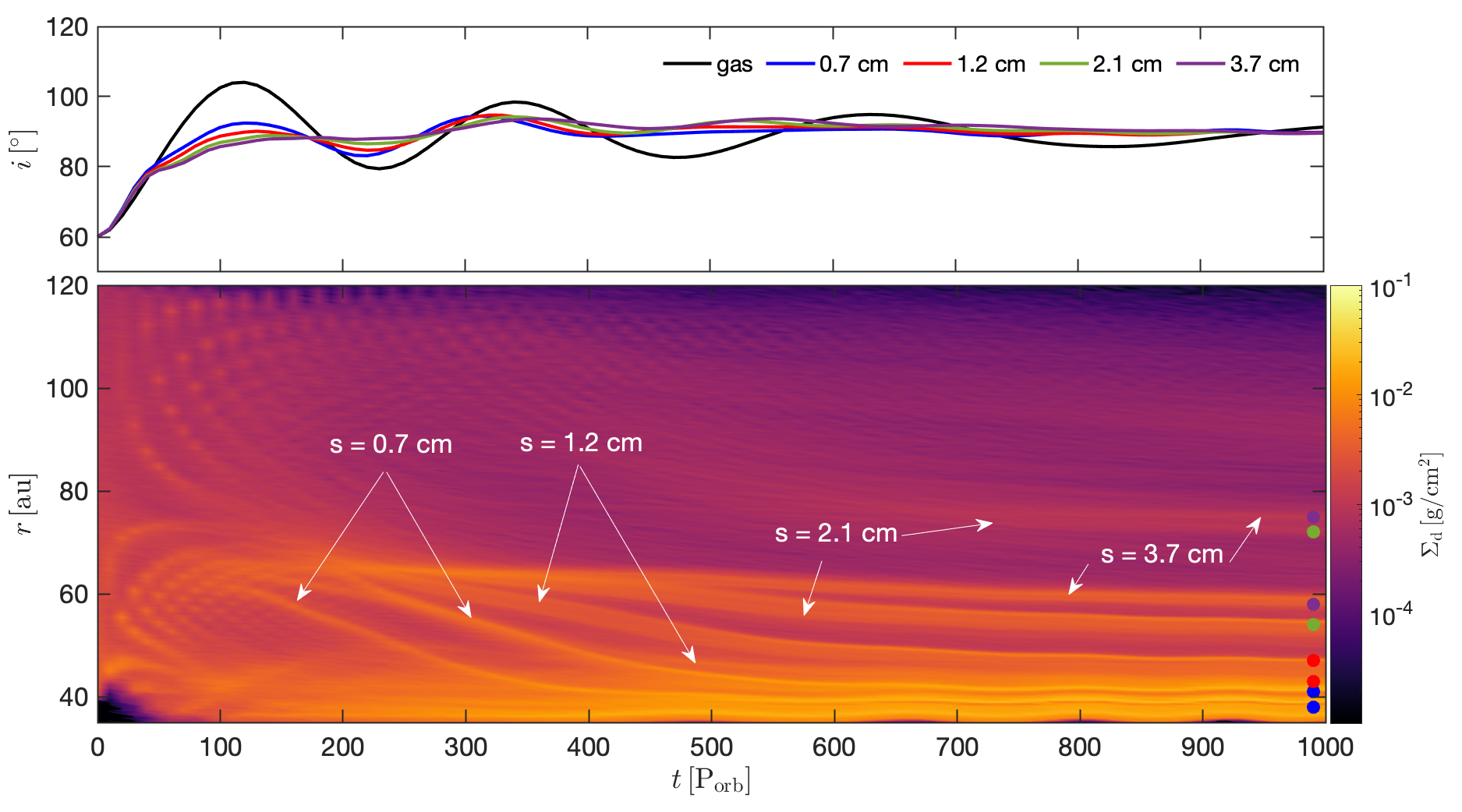}
\centering
\caption{ Top panel: the density weighted average of the disc inclination for the gas (black) and dust components,  $s = 0.7\, \rm cm$ (blue), $s = 1.2\, \rm cm$ (red), $s = 2.1\, \rm cm$ (green), and $s = 3.7\, \rm cm$ (purple).  Bottom panel: the azimuthally-averaged dust surface density, $\Sigma_{\rm d}$, as a function disk radius, $r$, and time in binary orbital periods, $\rm P_{orb}$. The colorbar denotes the dust surface density. The multiple dust rings are labeled by grain size, $ s$.  The colored dots indicate the location of the dust traffic jam at the end of the simulation for each dust size: $s = 0.7\, \rm cm$ (blue), $s = 1.2\, \rm cm$ (red), $s = 2.1\, \rm cm$ (green), and $s = 3.7\, \rm cm$ (purple).}
\label{fig::sigma}
\end{figure*}

\subsection{Radiative transfer calculations}
We use the results from our simulation as input to the Monte Carlo radiative transfer code {\sc mcfost} \citep{Pinte2006,Pinte2009}.  {\sc mcfost} is particularly well suited for particle-based numerical methods because it uses a Voronoi mesh rather than a cylindrical or spherical grid to generate a grid from the particles \citep{Camps2015}. Because the mesh follows the particle distribution, it does not require any interpolation \cite[e.g.,][]{Nealon2019}. Using a Voronoi mesh allows us to perform radiative transfer calculations on the complex geometry of a polar-aligned circumbinary disk. To address areas in the simulation with reduced numerical resolution, {\sc mcfost} incorporates a verification process that assesses the relationship between the cell size and the particle's smoothing length. If any dimension of a cell surpasses three times the smoothing length, {\sc mcfost} designates the area beyond three smoothing lengths of the particle within that cell as optically thin.

The binary star system will provide two sources of irradiation. Considering each star has a mass of $0.5\,\rm M_{\odot}$, we use a stellar spectrum and luminosity derived from a $3\, \rm Myr$ Siess isochrone \citep{Siess2000}: $T_{\rm eff} = 3758\, \rm K$, $L = 0.997\, \rm L_{\odot}$, and $R = 1.313\, \rm R_{\odot}$. We exclude any contribution arising from viscous heating. We employ $10^8$ photon packets for temperature calculations and determining monochromatic specific intensity. Dust optical properties are computed using Mie theory, assuming astrosilicate composition \citep{Draine1984}. We set the distance to $100\, \rm pc$ and the image size to $500\, \rm au \times 500\, au$ (equivalent to $5\, \rm arcsec \times 5\, arcsec$). Final images are generated using a ray-tracing method \citep{Pinte2009}.


\section{Hydrodynamical Results} 
\label{sec::Results} 
An initially misaligned gaseous circumbinary disk has the potential to undergo polar alignment, wherein the angular momentum of the disk precesses and ultimately aligns with the eccentricity vector of the binary system \cite[e.g.,][]{Aly2015,Smallwood2020}. Throughout this alignment process, oscillations in the disk tilt occur as a result of binary torque \citep{Smallwood2019}. As the disk approaches a nearly polar configuration, the amplitude of tilt oscillations diminishes.  The magnitude of the binary torque acting on the disk depends on both the tilt of the circumbinary disk and the eccentricity of the binary system. At a specific radius, the tidal torque becomes zero under two conditions: when the circumbinary disk is in a polar orientation and the binary eccentricity approaches $e_{\rm b} = 1$ \citep{Lubow2018}, or when the disk is in exactly a coplanar prograde or retrograde state \cite[][]{Nixon2013}. Given that the dust is still moderately coupled to the gas, during the alignment process, the gas will carry the dust into a polar configuration \citep{Smallwood2024a,Smallwood2024b}.

 During the alignment process, the gas and dust experience differential precession because the dust is decoupled from the gas. This differential precession results in the formation of multiple dust traffic jams. These occur at specific locations in the disc where the relative velocity between the gas and dust, accounting for pressure gradients, is locally minimised. In other words, dust traffic jams form when the velocity of the dust, projected onto the plane of the gas, matches that of the gas. It is important to note that this mechanism affects only the dust density profile, leaving the gas distribution unchanged. This differs from the conventional dust trap mechanism, where dust rings typically correspond to pressure maxima in the gas. In this case, multiple dust traffic jams may arise within the disc as a consequence of the differential precession of the gas and dust components \citep{Aly2020,Longarini2021, Aly2021, Smallwood2024a}.

\begin{figure*} 
\begin{center}
\includegraphics[width=\columnwidth]{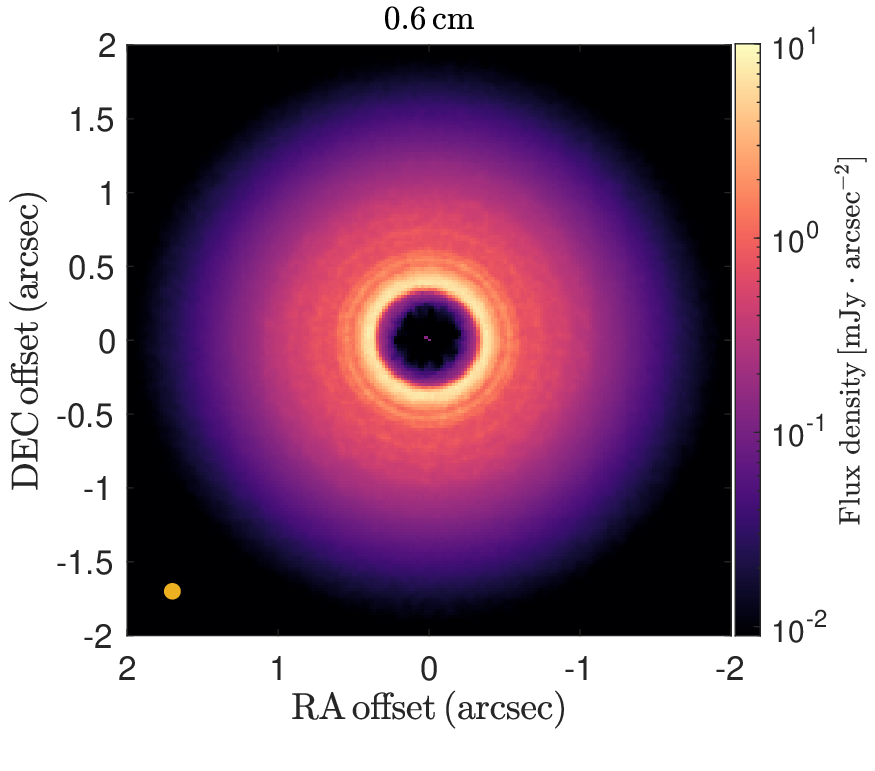}
\includegraphics[width=\columnwidth]{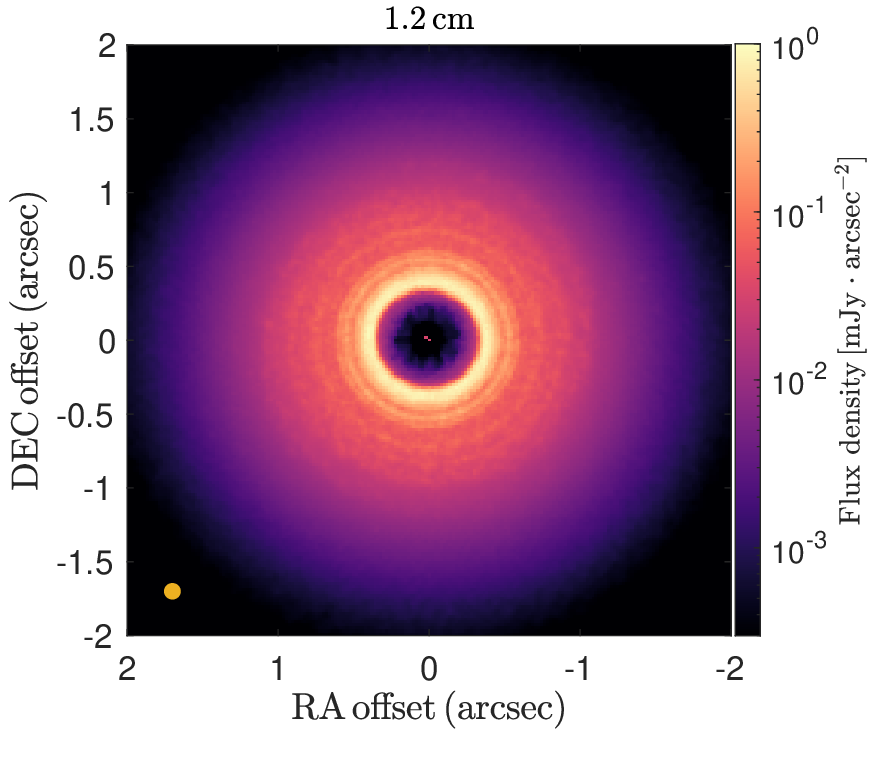}
\includegraphics[width=\columnwidth]{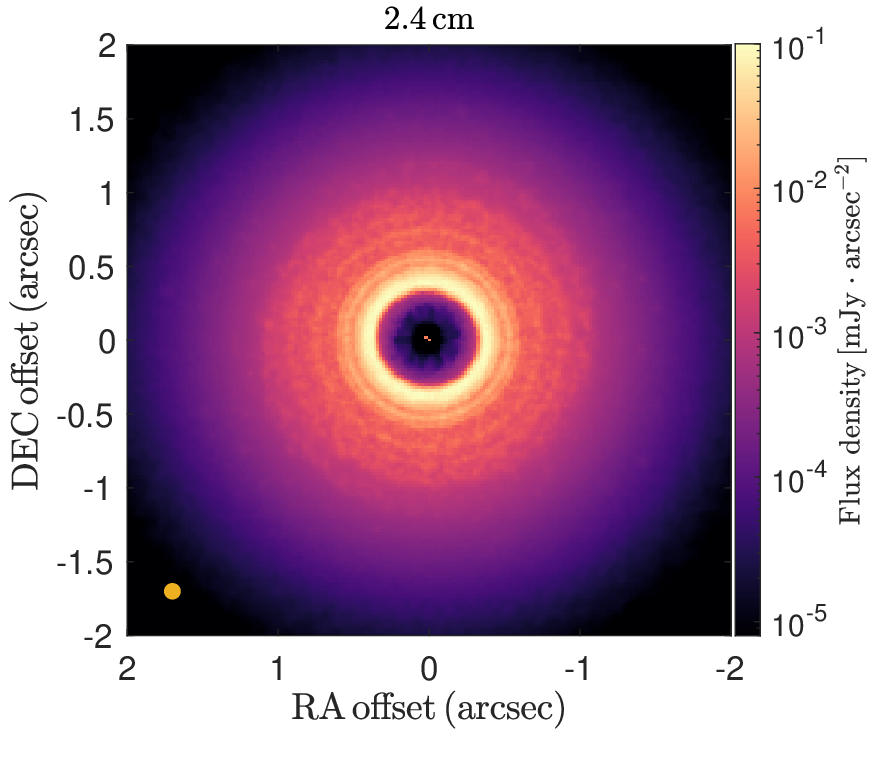}
\includegraphics[width=\columnwidth]{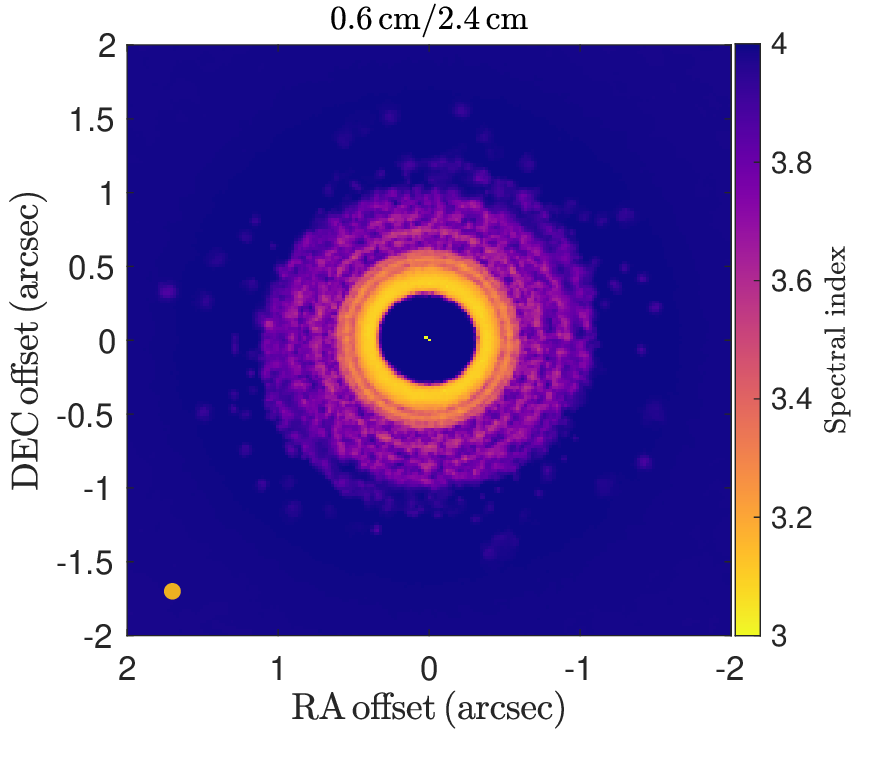}
    \includegraphics[width=\columnwidth]{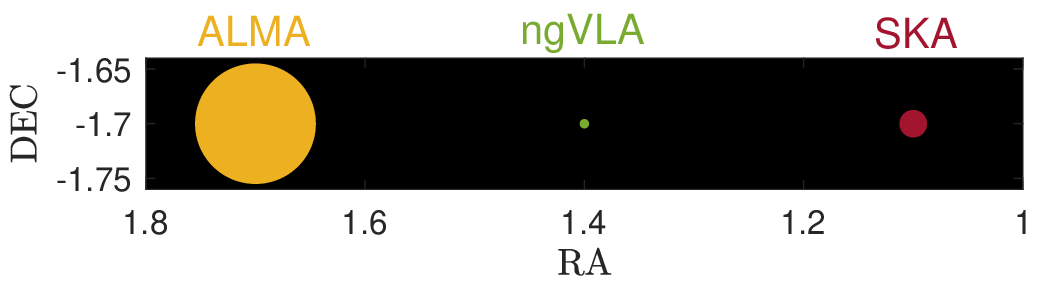}
\end{center}
\caption{The dust continuum emission at wavelengths $0.6\, \rm cm$ (top-left), $1.2\, \rm cm$ (top-right), and $2.4\, \rm cm$ (bottom-left).  Bottom-right panel: the spectral index, $\alpha$, of the dust continuum emission at wavelengths of $0.6\, \rm cm$ and $2.4\, \rm cm$, which reveals differences in the dust structure at different grain sizes. The emission from the two stars are also shown.  The estimated beam size for ALMA at $\sim 40, \rm GHz$ (yellow) is shown in the lower-left corner of each panel. The smaller subpanel below compares the beam sizes of ngVLA at $\sim 27, \rm GHz$ (green) and SKA at $\sim 11.9, \rm GHz$ (red) with that of ALMA.}
\label{fig::mcfost}
\end{figure*}


Figure~\ref{fig::splash} shows the structure of the circumbinary disk at $1000\, \rm P_{orb}$. At this time, the gas  and dust components are nearly aligned polar to the binary orbital plane,  as shown by the density-weighted tilt evolution in the upper panel of Fig.~\ref{fig::sigma}. However, we only show the dust component. The central image displays the circumbinary disk with all dust species visually represented. Each dust species is individually rendered and showcased in separate panels: $\rm s = 0.7\, cm$, $\rm s = 1.2\, cm$, $\rm s = 2.1\, cm$, and $\rm s = 3.7\, cm$. All panels view the disk in the $y$-$z$ plane (face-on to the polar disk). The color indicates the dust surface density. The differential precession between the gas and dust during polar alignment creates distinct dust rings contingent on grain size. The smaller the grain size, the faster they drift inward, which causes the dust ring to form and evolve nearer the disk's inner edge. As the grain size increases, the radial location of the dust rings occurs further out in the disk.


 The bottom panel in Fig.~\ref{fig::sigma} shows the azimuthally-averaged dust surface density as a function of disk radius and time. Multiple dust traffic jams are in the disk, each associated with a different grain size. Each dust traffic jam initially forms at a similar radial location ($r \sim 65\, \rm au$)
in the disk and drifts inward. The inward drift of the dust rings is faster for lower Stokes numbers (smaller grain size) since such particles experience more substantial gas drag \citep{Whipple1972,Weidenschilling1977}. Therefore, as the dust traffic jams drift inward, they split into multiple rings based on grain size. At $t\sim 700\, \rm P_{orb}$, two secondary dust traffic jams begin to form in the outer region of the disk, at $r\sim 75\, \rm au$, which are associated with the larger grain sizes, $\rm s = 2.1\, cm$ and $\rm s = 3.7\, cm$.

\section{Synthetic Observations} 
\label{sec::syn_obs} 

We compute the dust temperature structure using {\sc mcfost} and produce continuum images at wavelengths $0.6\, \rm cm$, $1.2\, \rm cm$, and $2.4\, \rm cm$ shown in Fig.~\ref{fig::mcfost}.  The estimated beam sizes are shown for ALMA at $\sim 40, \rm GHz$ (yellow), ngVLA at $\sim 27, \rm GHz$ (green), and SKA at $\sim 11.9, \rm GHz$ (red). For interferometric observations, the beam size is the angular resolution. In each continuum image, dust traffic jams can be seen within the disk.  The angular resolutions of ngVLA and SKA will be able to resolve the width of the dust traffic jams and the spiral arms.  The $0.6\, \rm cm$ continuum image traces the smaller grains, while the $2.4\, \rm cm$ continuum image traces the larger grains in the simulation. The main difference is the presence of dust spirals observed at $2.4\, \rm cm$ but not in the $0.6\, \rm cm$ continuum image. The larger grains are still misaligned to the binary at the end of the simulation \cite[see Figure 5 from][]{Smallwood2024a}, and thus cause the dust spirals at the longer wavelength synthetic observations. Due to a pressure bump near the inner edge of the disk caused by the binary \cite[see Appendix A in][]{Smallwood2024b} both small and large grains are trapped, causing the inner disk to be optically thick.

\begin{figure*}
\includegraphics[width=\columnwidth]{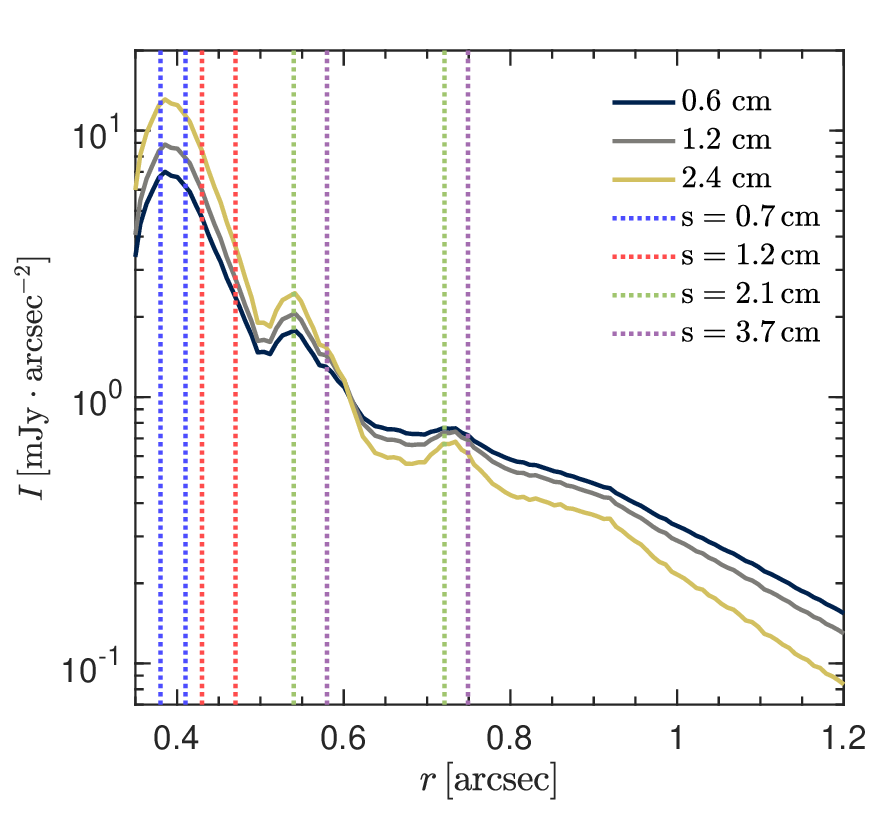}
\includegraphics[width=\columnwidth]{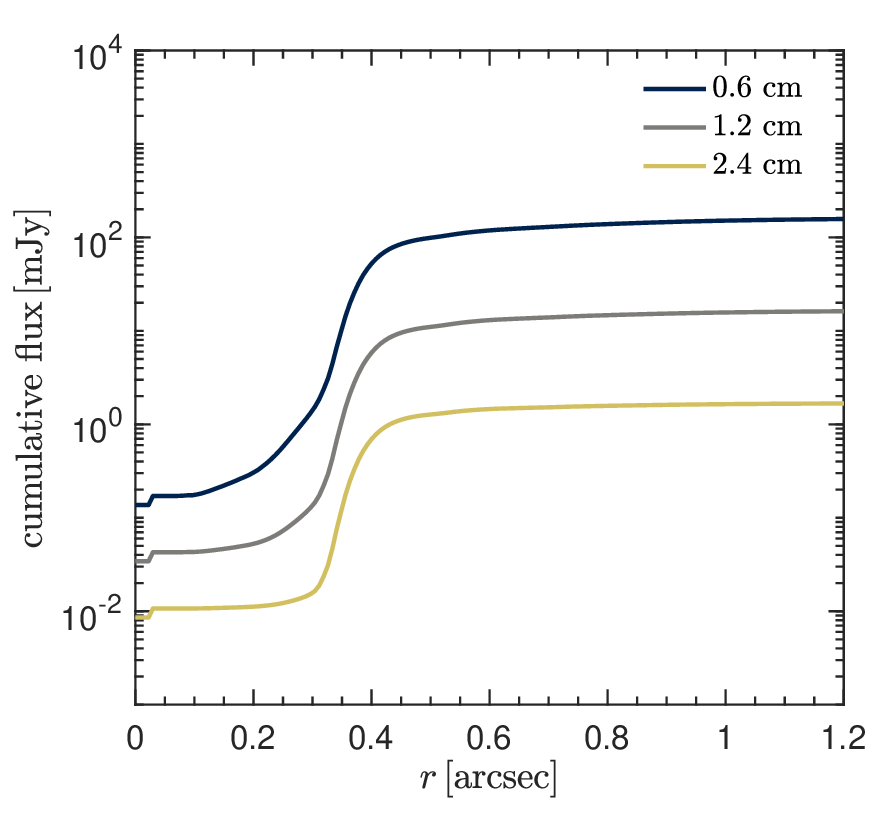}
\includegraphics[width=1.6\columnwidth]{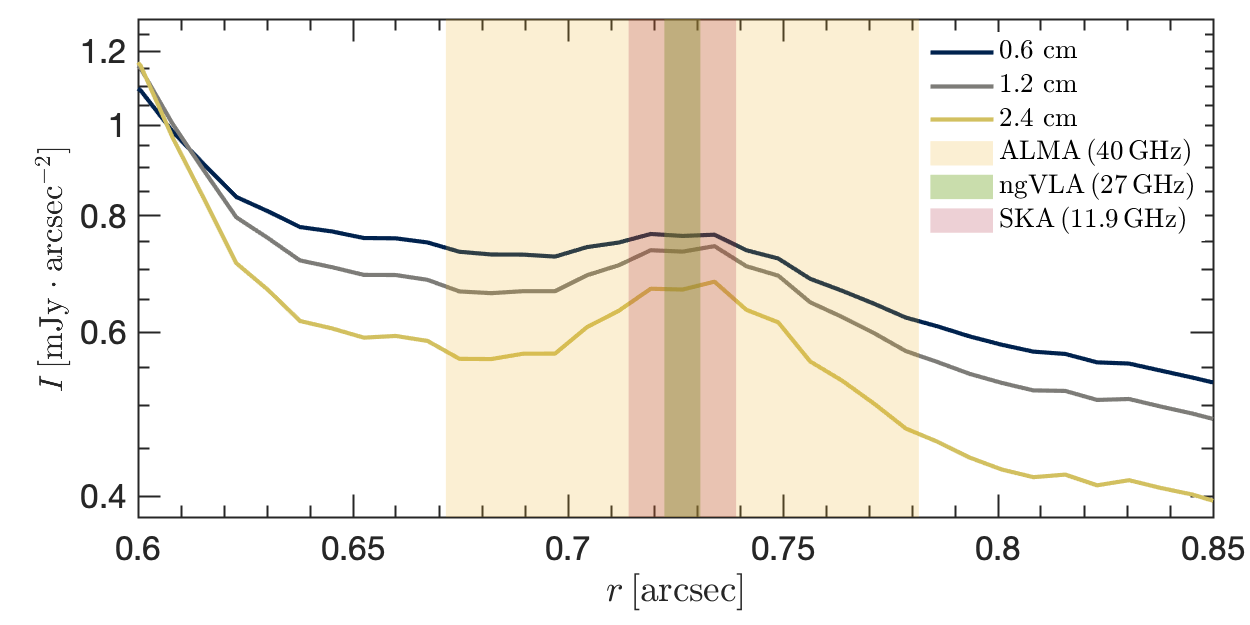}
\centering
\caption{ Upper left panel: the azimuthally-averaged flux density, $I$, as a function of distance from the image center from Fig.~\ref{fig::mcfost} for wavelengths, $0.6\, \rm cm$ (dark blue), $1.2\, \rm cm$ (grey), and $2.4\, \rm cm$ (yellow). The fluxes at $1.2\, \rm cm$ and $2.4\, \rm cm$ are scaled such that their initial flux profile (at $t=0\, \rm P_{orb}$) is equal to the initial flux profile of $0.6\, \rm cm$.  The vertical dashed lines correspond to the locations of the dust traffic jams  at $t = 1000\, \rm P_{orb}$ for each grain size, $\rm s = 0.7\, cm$ (blue), $\rm s = 1.2\, cm$ (red), $\rm s = 2.1\, cm$ (green), and $\rm s = 3.7\, cm$ (purple)  from the bottom panel in Fig.~\ref{fig::sigma}.  Upper right panel: the cumulative flux density as a function of distance from the image center from Fig.~\ref{fig::mcfost} for each wavelength.  Bottom panel: a close-up of the azimuthally-averaged flux density centered on the outer dust traffic jam, with the beam sizes of ALMA (at $40\, \rm GHz$), ngVLA (at $27\, \rm GHz$), and SKA (at $11.9\, \rm GHz$) overlaid as yellow, green, and red patches, respectively.}
\label{fig::flux}
\end{figure*}

The emission from dust grains of different sizes collectively contributes to the total flux. To compare the flux at two wavelengths, we compute the spectral index. The spectral index is given by
\begin{equation}
    \alpha = \frac{\log{(I_2/I_1})}{\log{(\nu_2/\nu_1})},
\end{equation}
where $I_1$ and $I_2$ are the flux densities at two different wavelengths with frequencies $\nu_1$ and $\nu_2$. The bottom right panel in Fig.~\ref{fig::mcfost} shows the spectral index between $0.6\, \rm cm$ and $2.4\, \rm cm$ wavelengths.
The temperature generally increases towards the inner regions of the disk due to higher stellar irradiation and possibly accretion heating. 
As a result, the inner optically thick regions tend to emit more strongly at shorter wavelengths, leading to a flatter or even inverted spectral index in those regions. This can manifest as a lower spectral index (closer to zero) in the inner regions compared to the outer, optically thinner parts of the disk, which typically exhibit a steeper spectral index (closer to the value expected for optically thin emission, \citealt{Dullemond2010,Williams2011}). Moreover, the presence of large dust grains in the optically thick regions can further alter the spectral index. Large grains can dominate the opacity, causing the spectral index to deviate from the standard values expected for smaller, interstellar medium-like grains. This effect is particularly pronounced at millimeter/centimeter wavelengths, where the scattering and absorption properties of large grains significantly influence the observed emission \citep{Ueda2023}.

The  upper left panel of Fig.~\ref{fig::flux} shows the azimuthally-averaged flux density, $I$, as a function of distance from the image center from Fig.~\ref{fig::mcfost} for wavelengths, $0.6\, \rm cm$, $1.2\, \rm cm$, and $2.4\, \rm cm$. The fluxes at $1.2\, \rm cm$ and $2.4\, \rm cm$ are scaled such that their initial flux profile (at $t=0\, \rm P_{orb}$) is equal to the initial flux profile of $0.6\, \rm cm$ for better comparison.  The vertical dashed lines represent the positions of the dust traffic jams at $t = 1000, \rm P_{orb}$ for each grain size: $\rm s = 0.7\, cm$ (blue), $\rm s = 1.2\, cm$ (red), $\rm s = 2.1\, cm$ (green), and $\rm s = 3.7\, cm$ (purple), as shown in the bottom panel of Fig.~\ref{fig::sigma}. The peak near $r\sim 0.4\, \rm arcsec$  arises from dust accumulation near the inner edge of the disk and the inner four dust traffic jams primarily associated with dust grains with $\rm St \sim 15$ and $25$. This dust accumulation near the inner edge \enquote{washes out} the signal from the dust traffic jams for the smaller grains, especially $s = 1.2\, \rm cm$.  There are two peaks in the flux profile occurring at $r \sim 0.55\, \rm arcsec$ and $r \sim 0.60\, \rm arcsec$, which corresponds to the larger grains with $s = 2.1\, \rm cm$ and $s = 3.7\, \rm cm$, respectively. Lastly, a peak at $r \sim 0.74\, \rm arcsec$ corresponds to secondary dust traffic jams comprised of larger dust grains, $s = 2.1\, \rm cm$ and $s = 3.7\, \rm cm$. This peak exhibits a more prominent presence at longer wavelengths (hardly seen at $0.6\, \rm cm$), giving evidence that larger dust grains are comprising the ring.

The  upper right panel of Fig.~\ref{fig::flux} shows the cumulative azimuthally-averaged flux density as a function of distance from the image center from Fig.~\ref{fig::mcfost}. The total flux of the source at wavelengths $0.6\, \rm cm$, $1.2\, \rm cm$, and $2.4\, \rm cm$ are $\sim 162\, \rm mJy$, $\sim 16.6\, \rm mJy$, and $\sim 1.7\, \rm mJy$, respectively. For $0.6\, \rm cm$, $50\%$, $68\%$, and $95\%$ of the total flux are within $0.39\, \rm arcsec$, $0.50\, \rm arcsec$, and $1.03\, \rm arcsec$, respectively. For $1.2\, \rm cm$, $50\%$, $75\%$, and $95\%$ of the total flux are within $0.38\, \rm arcsec$, $0.46\, \rm arcsec$, and $0.96\, \rm arcsec$, respectively. 
For $2.4\, \rm cm$, $50\%$, $75\%$, and $95\%$ of the total flux are within $0.36\, \rm arcsec$, $0.41\, \rm arcsec$, and $0.82\, \rm arcsec$, respectively. Given this information, the grain distribution is stratified, meaning smaller dust grains have drifted inward while the larger grains are still present in the outer regions of the disk (as expected).

The bottom panel of Fig.~\ref{fig::flux} shows a close-up of the azimuthally-averaged flux density centered on the outer dust traffic jam, with the beam sizes of ALMA at $40, \rm GHz$, ngVLA at $27, \rm GHz$, and SKA at $11.9, \rm GHz$ overlaid as yellow, green, and red patches, respectively. The angular resolutions at these frequencies are $0.11\, \rm arcsec$ for ALMA\footnote{\url{https://almascience.eso.org/proposing/sensitivity-calculator}}, $0.0084\, \rm arcsec$ for ngVLA\footnote{\url{https://ngect.nrao.edu/}}, and $0.025\, \rm arcsec$ for SKA\footnote{\url{https://sensitivity-calculator.skao.int/mid}}. The angular resolution of ALMA is too low to resolve the width of the dust traffic jam, but the angular resolutions of ngVLA and SKA are high enough to do so. Additionally, the resolution of ngVLA can be even higher than shown, depending on the array configuration \cite[e.g.,][]{Selina2018}.

\section{Discussion}
\label{sec::Discussion}
High angular resolution observations of disks at millimeter wavelengths, typically on scales of a few astronomical units, have become routine \citep{Andrews2018,Huang2018}. However, observations at centimeter wavelengths show a notable absence of comparable angular resolution. This discrepancy poses a significant obstacle to advancing our comprehension of planet formation and the process by which dust grains evolve from millimeter to centimeter sizes. Future telescope facilities will help bridge the gap between millimeter and centimeter observations.

Our hydrodynamical simulation considers relatively large dust grain sizes ranging from $0.7\, \rm cm$ to $3.7\, \rm cm$. This range is explicitly chosen to accurately model the formation of dust traffic jams through differential precession, necessitating a two-fluid hydrodynamical model. To make observational predictions from our synthetic observations displayed in Figs.~\ref{fig::mcfost} and~\ref{fig::flux}, we invoke the use of upcoming Square Kilometre Array (SKA) and next-generation Very Large Array (ngVLA) observations, which will be able to probe cm-sized dust grains. The VLA can conduct lower resolution observations at cm-wavelengths from the $\rm Disks@EVLA$ Program \cite[e.g.,][]{Perez2015,Guidi2016,Tazzari2016}. Recently, \cite{Carrasci-Gonzalez2019} used ALMA and VLA to observe the HL Tau Disk at a range of wavelengths, from $0.8\, \rm mm$ to $1\, \rm cm$. Characterizing the distribution of dust grain sizes in protoplanetary disks will increase our understanding of planetesimal growth. However, current instrumentation lacks the necessary spatial resolution and sensitivity for characterization.


SKA will soon provide higher-resolution interferometric observations at cm-wavelengths \cite[e.g.,][]{Braun2015}. The mid-frequency array, SKA1-MID, will have maximum baselines of $150\, \rm km$, providing extremely high-resolution observations at frequencies between $350\, \rm MHz$ and $15.3\, \rm GHz$. The proposed Band5b will have a frequency range of $8.3-15\, \rm GHz$ and a wavelength range of $3.6-2.0\, \rm cm$ with angular resolutions down to $0.04\, \rm arcsec$ \citep{Braun2019}. The next-generation VLA (ngVLA) is an interferometric array that significantly enhances sensitivity and spatial resolution over the Janksy VLA and ALMA at the same wavelengths \citep{Murphy2018}. Operating within frequencies ranging from $1.2\, \rm GHz$ (25 cm) to $116\, \rm GHz$ (2.6 mm), the ngVLA will serve as a crucial bridge between ALMA and the forthcoming SKA, addressing a critical gap in existing observational capabilities.

\begin{figure*} 
\includegraphics[width=0.69\columnwidth]{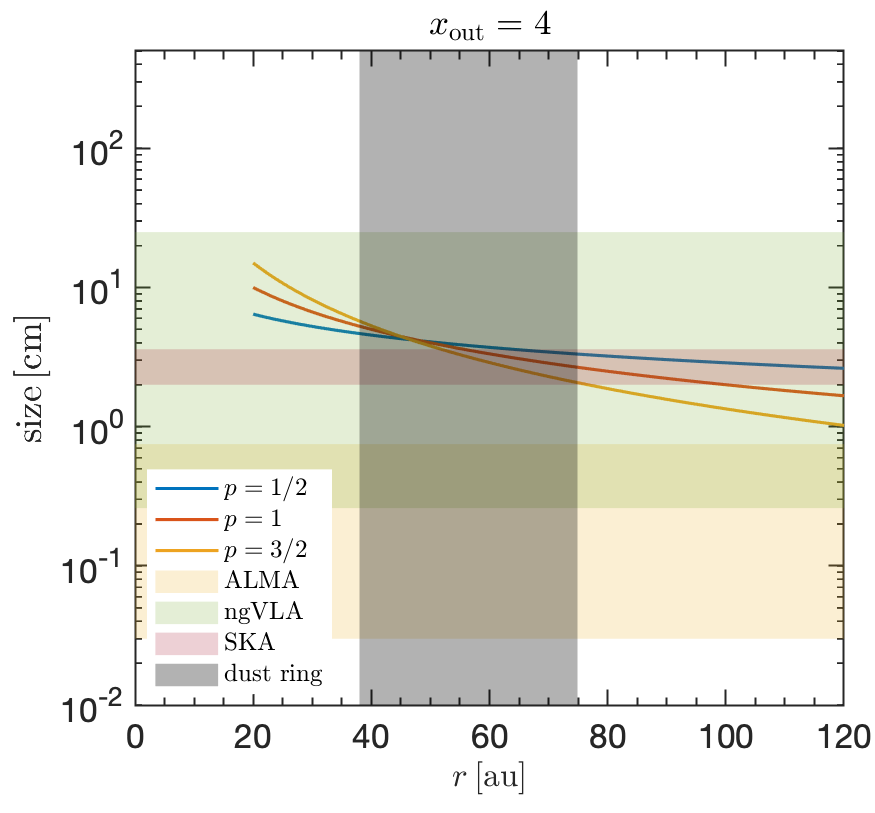}
\includegraphics[width=0.69\columnwidth]{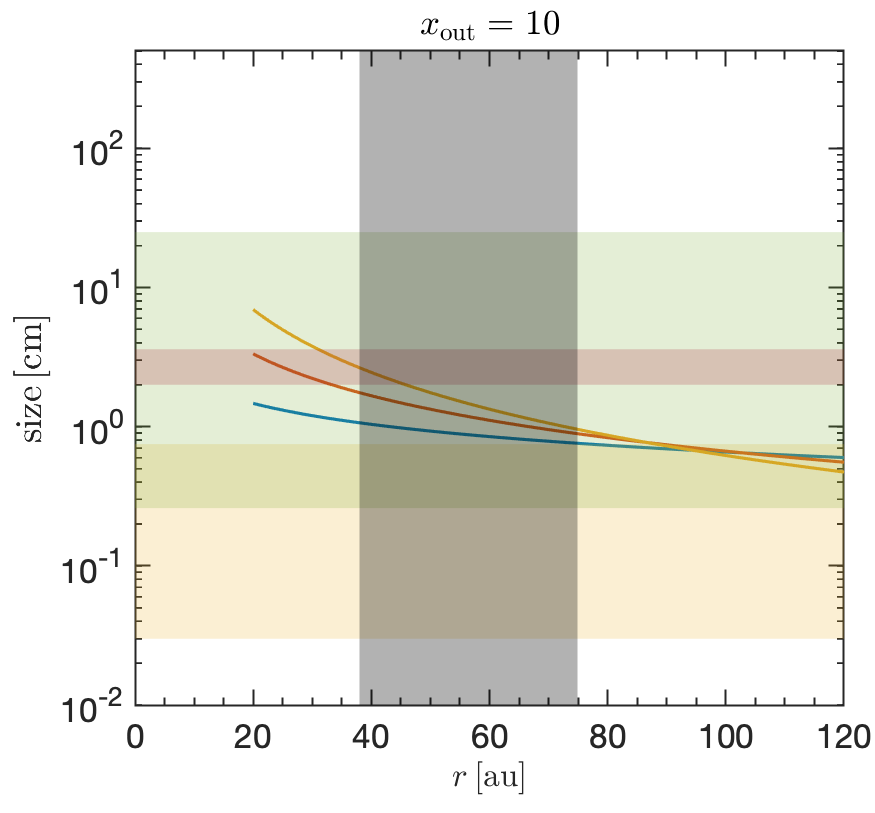}
\includegraphics[width=0.69\columnwidth]{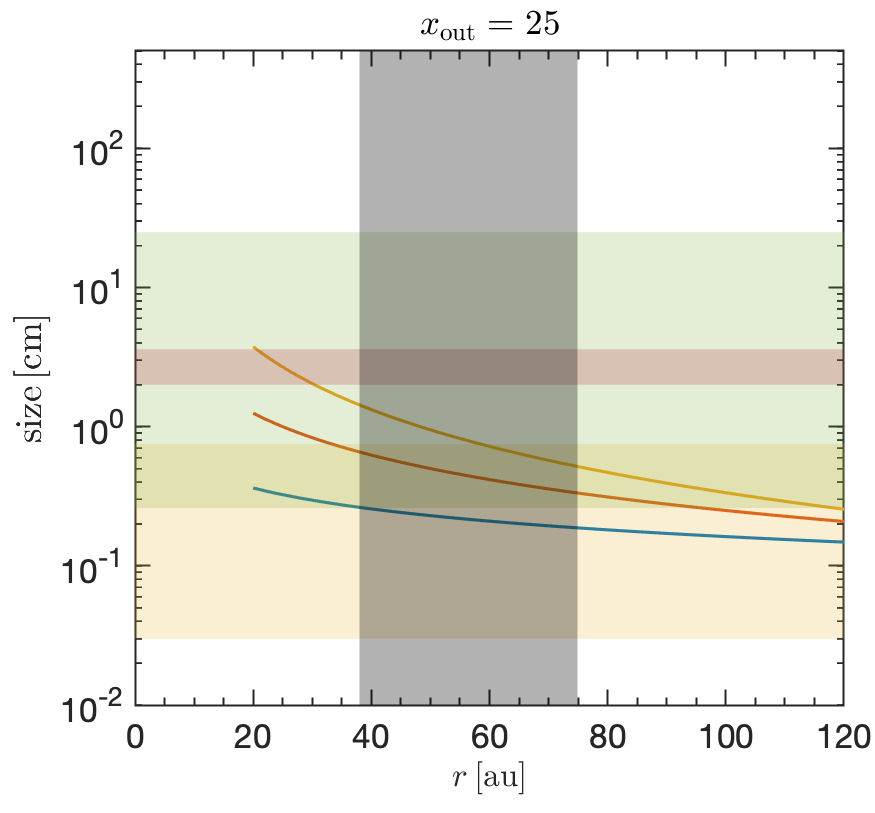}
\includegraphics[width=0.69\columnwidth]{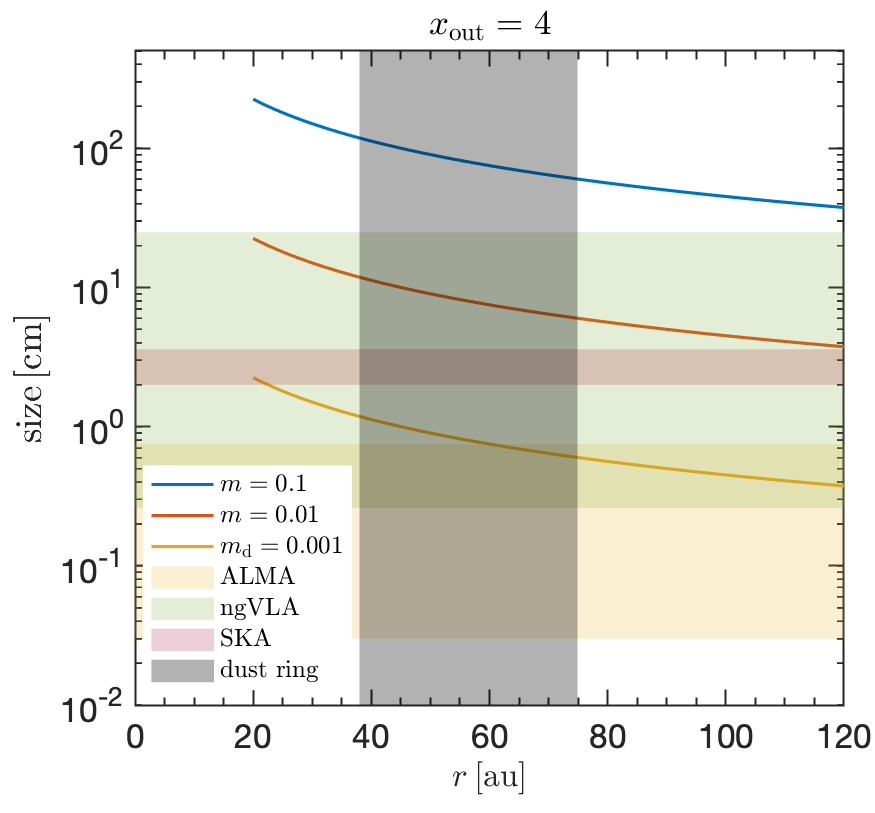}
\includegraphics[width=0.69\columnwidth]{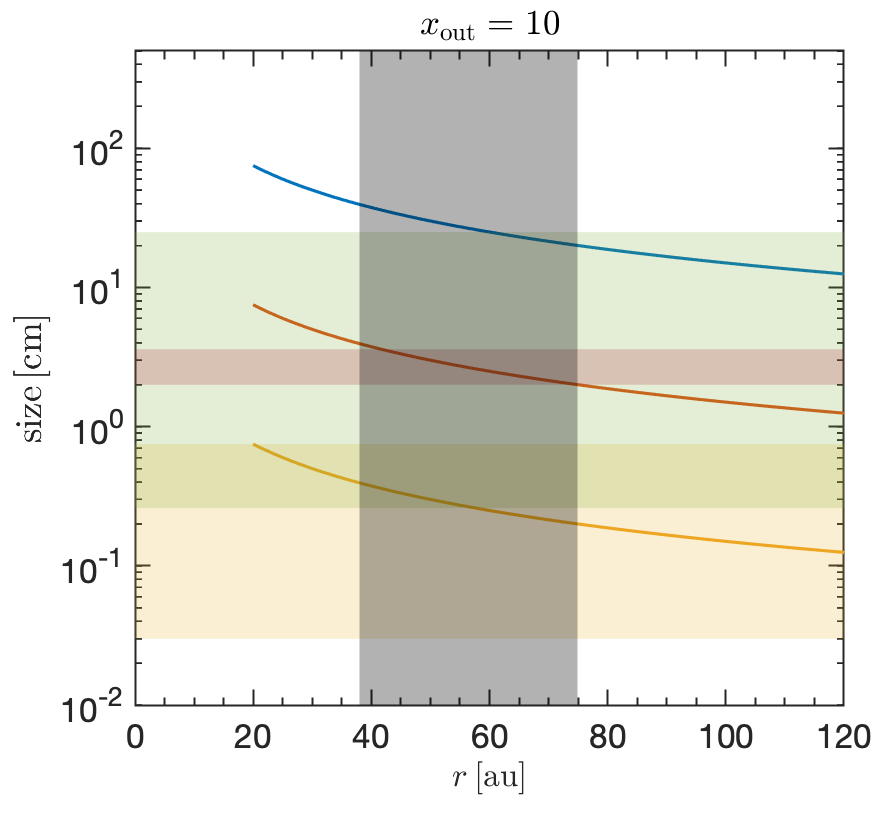}
\includegraphics[width=0.69\columnwidth]{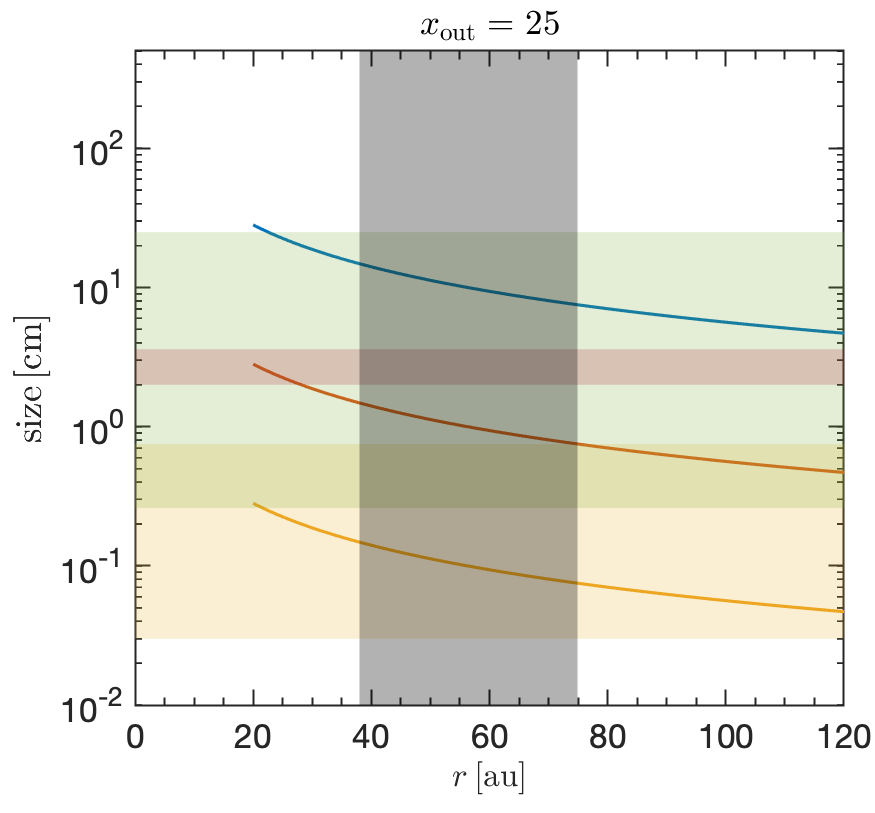}
\centering
\caption{The size of $\rm St \sim 1$ dust particles as a function of radius $r$ evaluated for different values of the surface density radial profile $p$, with the disk mass $M_d = 0.01$ (top row), and for different values of $M_d$, with $p = 1$ (bottom row). Each column represents a different value of $x_{\rm out} = r_{\rm out}/r_{\rm in}$. The yellow, green, and red shaded regions represents the observational window of ALMA $[0.03\, \rm cm;\, 0.75\, \rm cm ]$, ngVLA $[0.26\, \rm cm;\, 25\, \rm cm ]$, and SKA $[2\, \rm cm;\, 3.6\, \rm cm ]$, respectively.  We highlight the region in the disc where we expect to observe the dust traffic jams (black shaded region) based on the results of our simulations from Fig.~\ref{fig::sigma}.}
\label{fig::size}
\end{figure*}

From this and previous studies, dust rings are likely to form for moderately coupled dust particles $\rm St  \gtrsim 1$ \cite[e.g.,][]{Longarini2021,Aly2021,Smallwood2024a,Smallwood2024b}. To further study the possible observability of these dust traffic jams, we employ a simple analytical approach detailed in \cite{Longarini2021}.  Although our hydrodynamical simulations modeled dust grains with $\rm St \gtrsim 15$, in our toy model we can assume that $\rm St \sim 1$ represents a lower limit for the formation of dust rings.  The effect of dust radial drift is larger for when $\rm St \simeq 1$, given by
\begin{equation}
    {\rm St} = \frac{\pi \rho_0 s}{2 \Sigma_{\rm g}} \simeq 1.
    \label{eq:stokes}
\end{equation}
Equation~(\ref{eq:stokes}) can be rewritten in terms of the disk parameters,
\begin{equation}
    s(r) = \frac{2}{\pi} \frac{\Sigma_{\rm g}(r)}{\rho_0} = \frac{2}{\pi \rho_0} \frac{M_d (2-p)}{2\pi r^2_{\rm in}} \frac{1}{x^{2-p}_{\rm out} -1} \bigg( \frac{r}{r_{\rm in}}\bigg)^{-p},
        \label{eq:size}
\end{equation}
where $M_d$ denotes the disk mass and $x_{\rm out} = r_{\rm out}/r_{\rm in}$. Additionally, we have adopted a pure power-law density profile for the gas. From equation~(\ref{eq:size}), we can calculate the size of $\rm St \sim 1$ dust particles at any radius $r$, depending on the disk parameters. The top row Figure~\ref{fig::size} shows the  size of $\rm St \sim 1$ dust particles evaluated for different values of the density parameter $p$, with $M_d = 0.01$ (top row), and for different values of the disk mass $M_d$, with $p = 1$ (bottom row). The yellow, green, and red shaded regions represents the observational window of ALMA, ngVLA, and SKA, respectively. We highlight the region in the disc where we expect to observe the dust traffic jams (black shaded region) based on the results of our simulations from Fig.~\ref{fig::sigma}. Each column represents a different value of $x_{\rm out} = r_{\rm out}/r_{\rm in}$, with $x_{\rm out} = 4$ corresponding to our SPH simulation. From Fig.~\ref{fig::size}, there are several combinations of $p$, $M_d$, and $x_{\rm out}$ whereby the dust traffic jams may be detectable for ngVLA, SKA, and even with ALMA.  However, the angular resolution of ALMA may not be enough to resolve the width of the dust traffic jams (refer back to the bottom panel in Fig.~\ref{fig::flux}).

\cite{Ilee2020} used hydrodynamical simulations coupled with radiative transfer to predict the distribution and emission of $\sim1\, \rm cm$ dust grains using simulated SKA observations. They found that the SKA can produce high-resolution observations of cm dust emission in disks given a total integration time of 100 hours.  However, the disk radial structure may be observed with medium resolution with an integration time of 10 hours by azimuthally averaging the image plane. \cite{Wu2024} used a combination of ALMA, SKA, and ngVLA to produce synthetic observations of protoplanetary disks at sub-cm/cm wavelengths. Based on their results, future SKA and ngVLA observations will complement pre-existing ALMA observations by probing cm-sized grains. Therefore, future observations using SKA and ngVLA may be used to observe dust traffic jams in initially misaligned circumbinary disks. 

 As a circumbinary disc evolves to polar alignment, strong warps are produced as well as spiral generated by the binary potential. At the same time, dust traffic jams are produced by the differential precession between the gas and dust \citep{Aly2020,Longarini2021,Aly2021,Smallwood2024a}, producing ring-like structures.  Ring structures may also be produced by magnetic field concentrations in MHD zonal flows \citep{Suriano2018} or perturbations from low-mass planets \citep{Zhu2011,Rosotti2016}. However, the locations of these dust rings are independent of grain size, unlike the dust traffic jams formed in misaligned circumbinary discs. Additionally, a vortex can produce a ring–cavity substructure morphology with a pronounced asymmetric feature \citep{Baruteau2016}, and  spiral wave perturbations may be produced by the global gravitational instability driven by remnant envelope infall \citep{Lesur2015} or tidal interactions with a massive (external) planetary companion \citep{Dong2015}. 

 Based on our synthetic observations, larger grain sizes require more time to achieve a polar configuration. As a result, dust spirals are often produced during polar alignment due to the binary potential. However, observations of the polar disk around HD 98800B at 1.3 mm revealed no detection of such dust spirals \citep{Kennedy2019}. One possibility is that these grains drifted inward and aligned polar, leaving no visible extended spiral arms (see upper left panel in Fig.~\ref{fig::mcfost}). Alternatively, the absence of observed spirals could be due to the resolution limitations of the ALMA beam, which may not be sufficient to resolve the spiral arms (see bottom panel in Fig.~\ref{fig::flux}).


\section{Conclusion}
\label{sec::Conlusion}
We conducted a hydrodynamical simulation of an initially misaligned circumbinary disk undergoing polar alignment with multiple dust species. Multiple dust traffic jams are produced within the disk due to differential precession between the gas and dust components. The ultimate radial location of the dust traffic jams depends on the Stokes number of the grains. We computed the dust temperature structure using the radiative transfer code \textsc{mcfost} to produce continuum images at $0.6\, \rm cm$, $1.2\, \rm cm$, and $2.4\, \rm cm$. Multiple peaks emerge in the azimuthally-averaged flux density analysis, representing the dust traffic jams. As traffic jams are dependent on St, multiwavelength observations become an important predictive tool to identify this phenomenon.  The angular resolution of future SKA and ngVLA observations will be capable of detecting centimeter-sized grains in protoplanetary disks and resolving the widths of dust traffic jams. As a result, dust traffic jams caused by the differential precession of gas and dust in misaligned circumbinary disks will be key targets for extended wavelength observations. 

\section*{Acknowledgments}
 We thank the anonymous referee for helpful revisions that improved the quality of the manuscript. JLS acknowledges funding from the ASIAA Distinguished Postdoctoral Fellowship and the Taiwan Foundation for the Advancement of Outstanding Scholarship.  RN acknowledges funding from UKRI/EPSRC through a Stephen Hawking Fellowship (EP/T017287/1). CL acknowledges funding from the European Union’s Horizon 2020 research and innovation programme under the Marie Skłodowska-Curie grant agreement No 823823 (RISE DUSTBUSTERS project). HA acknowledges funding from the European Research Council (ERC) under the European Union’s Horizon 2020 research and innovation programme (grant agreement No 101054502).  MKL is supported by the National Science and Technology Council
(grants 111-2112-M-001-062-, 112-2112-M-001-064-, 111-2124-M-002-013-, 112-2124-M-002 -003-) and an Academia Sinica Career Development Award (AS-CDA110-M06).

%

\vspace{5mm}


\software{\sc{phantom} \citep{Price2018}, {\sc mcfost} \citep{Pinte2006,Pinte2009}.}





\bibliography{ref}
\bibliographystyle{aasjournal}



\end{document}